# Design and Analysis of a Metamaterial-Inspired Absorber for data rate in 52% RF-to-DC conversion Efficiency Dual-band SWIPT system

ZhengChen Dong[1], Xin,Jiang[2], Ramesh K. Pokharel[2], Adel Barakat
Graduate School of Information Science and Electrical Engineering, Kyushu University, Japan
[1]dong.zhengchen.204@s.kyushu-u.ac.jp, [2]pokharel@ed.kyushu-u.ac.jp

*Abstract*—This paper proposes a novel metamaterial-inspired absorber designed to enhance the data rate in 52% RF to DC conversion simultaneous wireless information and power transfer system (SWIPT) through biological tissue. The proposed absorber includes split-ring resonators(SRRs) and demonstrates significant permeability characteristics, with both the real and imaginary parts being negative and close to -1. It also improves isolation by around 5dB in a WPT distance of 9mm. A 5mm thick phantom is used for biological tissue in this study. Experimental results exhibits that the SWIPT systems including a rectifier that converts 52% RF to DC efficiency in a WPT distance of 9mm embedding this absorber between power and signal ports at Tx side results in a 5dB improvement in isolation performance. By using proposed absorber, it enables a 7MB/s improvement of data rate and allows signals to be transmitted with 5dBm weaker power than without absorber SWIPT system.

## I. INTRODUCTION

Many portable, implantable and wearable electronic devices[1],[2] has been used today, which causes an unprecedented significance of compact and efficient wireless charging systems. Concretely, Simultaneous wireless information and power transfer (SWIPT) systems can offer better energy harvesting and rational utilization as well[3][4].

Moreover, as introduced in [5],[6], a 2-port SWIPT system using the load shift key(LSK) modulation technique which can not only transfer power but transmit data as well was introduced. However, this SWIPT system requires a modulator and demodulator at Tx and Rx, which causes a poor isolation between two channels.

As showed in Fig.1(a), To solve these problems, a compact 4-port simultaneous wireless information and power transfer system(SWIPT) was proposed[7]. This system enables the SWIPT system to transfer power and transmit signal with stronger isolation and higher data rate than previous SWIPT system in lower frequency band, which also means the power band. In terms of isolation and phase noise performance, metamaterials have been extensively studied for their applications as not only isolators[8],[9] but circulators[10][11],absorbers[12]as well, in the far field. In contrast, for inductive coupling(near field) WPT and SWIPT systems, the metamaterial is used to enhance power transfer efficiency(PTE) [13]-[19], performance in lateral misalignment[20][21], to degrade specific absorption rate(SAR)[22], and so on[23].

However, the SWIPT system in [7] also remains a critical problem that the introduced system need to improve the isolation in higher frequency band. Therefore, in this study, inspired by metamaterial[8][24], to go further, we proposed a metamaterial-inspired absorber using two split-ring resonators(SRRs) to enhance the data rate in higher band showed in Fig.1(b) .two split ring resonator(SRRs) are etched in the middle of two ports in Tx. The power is transferred from Port 1 to Port 3 in the lower frequency band and the signals are transmitted from Port 2 to Port 4 in the higher frequency band. The experiment includes two human body equivalent phantom sandwiching the Rx in the middle. The proposed absorber is embedding in Tx side and the rectifier is connected with Rx in power port. The results of bit error rate(BER) and S parameters indicate that the proposed absorber improve the isolation and data rate in higher frequency band, which allows the SWIPT system can transmit the signal with weake power than conventional SWIPT system. Additionally, the rectifier in Rx side keep higher Rf-to-Dc conversion efficiency in 52% and is more compact than conventional SWIPT system used in biological tissues.

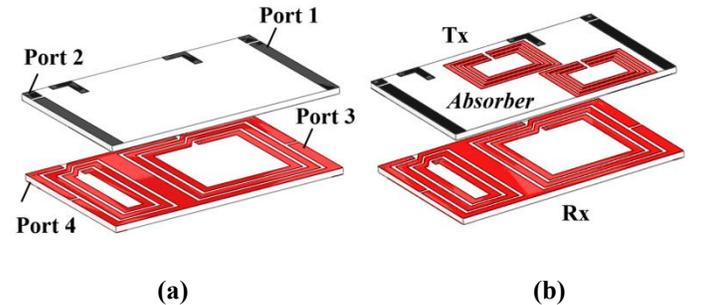

Fig.1(a) 4-Port conventional SWIPT system.(b) proposed 4-Port SWIPT system.

## II. MATH AND THEORETICAL ANALYSIS OF PROPOSED SWIPT SYSTEM

Metamaterial-based structures have widely been employed in far-field antennas [24] or meta-antennas [10]. However, there are few applications of metamaterial-inspired absorber has been used on near field system. This part proposes a metamaterial -inspired absorber in the near-field SWIPT system.

### A. Equivalent circuit and CRLH-TL theory

Fig.2 illustrates the equivalent circuit of metamaterial's unit cell. The distinctive characteristics of these metamaterials arise from parasitic capacitance and mutual inductance between each cell.

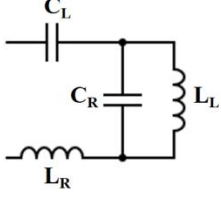

Fig.2 equivalent circuit of metamaterial unit cell.

The properties of the proposed metamaterial are affected by several factors: the capacitance of the line($C_L$), the inductance of the line($L_L$), as well as the parasitic capacitance($C_R$) and mutual inductance($L_R$). These components collectively influence the overall behavior and performance of the metamaterial.

The proposed metamaterial has two resonance frequencies, the series resonance frequency($\omega_{se}$) and the shunt resonance frequency($\omega_{sh}$). Its characteristic impedance $Z_c$ can be determined based on these frequencies accordingly[25],[26].Specifically, at the series resonance frequency, the impedance is minimized, and at the shunt resonance frequency, the impedance is maximized. These resonances are crucial for defining the behavior of the metamaterial and enhancing the isolation performance in SWIPT systems by blocking unwanted signal propagation and reducing mutual inductance.

$$Z_c = \sqrt{\frac{L_L}{C_L}} \times \sqrt{\frac{\left(\frac{\omega}{\omega_{se}}\right)^2 - 1}{\left(\frac{\omega}{\omega_{sh}}\right)^2 - 1}} \quad (1)$$

$$\omega_{se} = \left(\sqrt{L_R C_L}\right)^{-1} \quad (2)$$

$$\omega_{sh} = \left(\sqrt{L_L C_R}\right)^{-1} \quad (3)$$

From the equation above, in higher band, when ω close to $\omega_{sh}$, the absorber effectively acts as an open circuit. As a result, the wave propagation from Port 2 to Port 1(also Port 4 to Port 3). Consequently, the mutual inductance $M_{12}$(Mutual induction of Port 1 and Port 2), $M_{34}$(Mutual induction of Port 3 and Port 4). Simultaneously, $M_{14}$ and $M_{23}$ will also reduce by the influence of less energy gathering, which can improve the isolation performance.

Moreover, when ω equals to $\omega_{sh}$.The magnetic wave energy is totally absorbed and stored within the SRRs. In this case, no power can be transferred from Tx to Rx. However, when ω close to $\omega_{se}$. The absorber becomes a short circuit because $Z_c$ is nearly 0. In this case, more magnetic waves pass through the absorber, which will increase $M_{12}$, $M_{34}$, also $M_{23}$ and $M_{14}$.

### B. Refraction and permeability analysis

In this section, we focus on index of refraction and analysis the permeability of proposed absorber.

According to [27], the transmission coefficient of the H-field is:

$$T = tt'e^{ik'_x d} \quad (4)$$

Where t and t′ are the transmission when the wave is on and inside Tx, respectively. And $k_x'$ is the wave number of wave in x-direction. The proposed absorber is on the bottom of Tx and then there is:

$$t = \frac{2\mu k_x}{\mu k_x + k'_x}, \quad t' = \frac{2k_x}{\mu k_x + k'_x} \quad (5)$$

Where $k_x$ is the wave number of the incident wave from free space. According to [27], we can figure out the transmission inside Tx:

$$T_{Tx} = \frac{tt'e^{ik'_x d}}{1 - r'e^{2ik'_x d}} \quad (6)$$

r' means the reflection of wave inside Tx and r' is:

$$r' = \frac{k'_x - \mu k_x}{k'_x + \mu k_x} \quad (7)$$

When $\mu \rightarrow -1$, the transmission of Tx is:

$$\lim_{\mu \rightarrow -1} T_{Tx} = e^{-ik_x d} \quad (8)$$

From (8) and (4), it is evident that when the permeability($\mu$) approaches -1,the wave propagating in the x-direction reverse with the assistance of the proposed metamaterial, which results in a refractive index that is nearly -1.

As illustrated in Fig.3, Fig.3(a) shows a 3-dimensional view of SWIPT model. Fig.3(b) shows the propagation of incident wave and refracted wave of H-field in XZ-plane. When using an absorber with negative permeability properties, it undergoes reverse refraction.

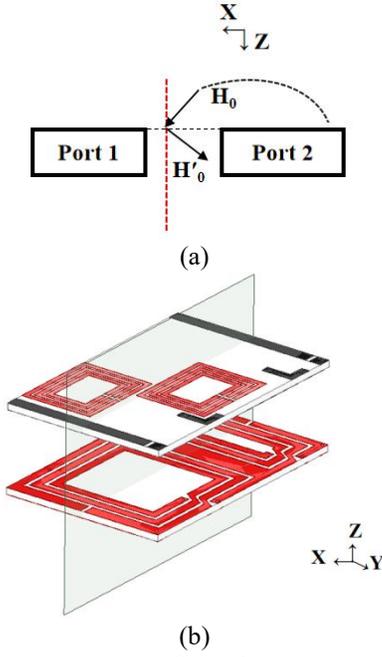

(a)

(b)

Fig.3(a)The propagation of incident wave and refracted wave of H-field in XZ-plane.(b) SWIPT system 3-D model.

*C. EM Simulation of proposed absorber*

The model of proposed absorber has been showed in Fig.4 and the dimensions specified in Table I. In this study, we use high frequency simulation software(HFSS) to assess the performance of the proposed absorber. Following the method in [28], the optimization method that make the permeability of proposed absorber unit cell -1.

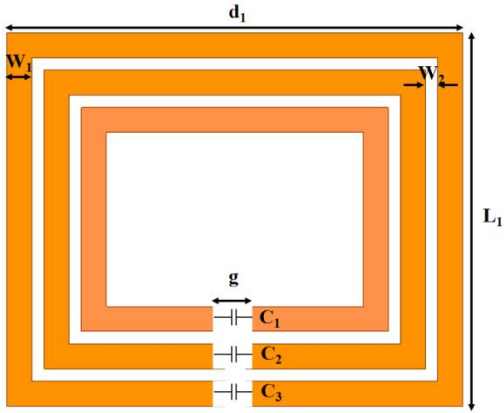

Fig.4 Model of proposed absorber.

Table I
DIMENSION OF PROPOSED ABSORBER

| $d_1$ | $W_1$ | $W_2$ | $L_1$ | g | $C_1$ | $C_2$ | $C_3$ |
|---|---|---|---|---|---|---|---|
| 11mm | 0.6mm | 0.3mm | 9mm | 0.5mm | 510pF | 620pF | 620pF |

The Tx and Rx components are produced by printing 0.017mm thick copper onto a substrate of Rogers3003. This substrate possess a dielectric constant of 3, a thickness of 0.762mm and a loss tangent of 0.001.

As depicted in Fig.5, Fig.5 shows the estimated permeability of the proposed absorber.

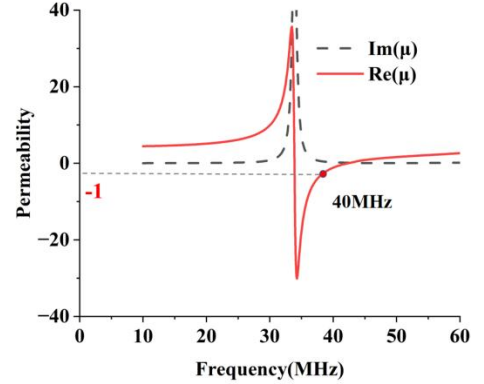

Fig.5.estimated permeability of the proposed absorber.The higher band of WPT system 40MHz and permeability at 40MHz is almost -1.

At the operating frequency of 40MHz, the proposed absorber exhibits a real permeability value close to -1. Simultaneously, the imaginary part of the permeability nears 0, ensuring minimal magnetic losses. This facilitates power transfer to Rx from Port 1 to Port3.

Moreover, Fig.6 shows H-field distribution within the SWIPT system at 40MHz, both with and without the proposed absorber, in XY plane. From Fig.6(a) and Fig.6(b), the proposed absorber reduce the H -field leakage from Port 2 to Port 1, which also means that more magnetic energy can be stored in Port 2. As mentioned in Section II A(3), reciprocity ensures that the same effect will occur in reverse scenarios.

Simultaneously, From Fig.6(c) and Fig.6(d), the mutual induction of both Port 1 and Port 4($M_{14}$) and Port 1 and Port2($M_{12}$) is reduced, which means that the isolation between Port 1 and Port 4 is enhanced, and Port 1 and Port 2 as well,which can enhance the PTE from Port 1 to Port3.

Due to the isolation decreased in 40MHz higher band, the data rate can be improve and its also needs less power with the help of less magnetic leakage.

In conclusion, the simulated results show great agreement with the theories presented in the previous subsection. For the proposed WPT system, the WPT frequency is supposed to be located where permeability is near -1 in the real part and near 0 in the imaginary part, respectively.

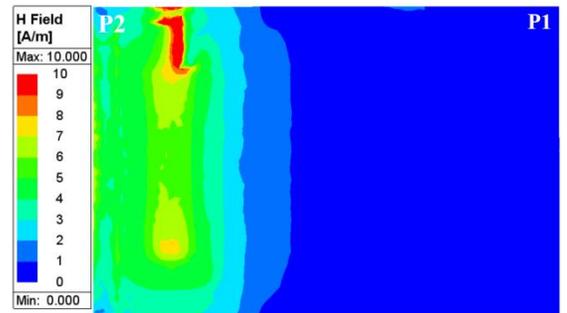

(a)

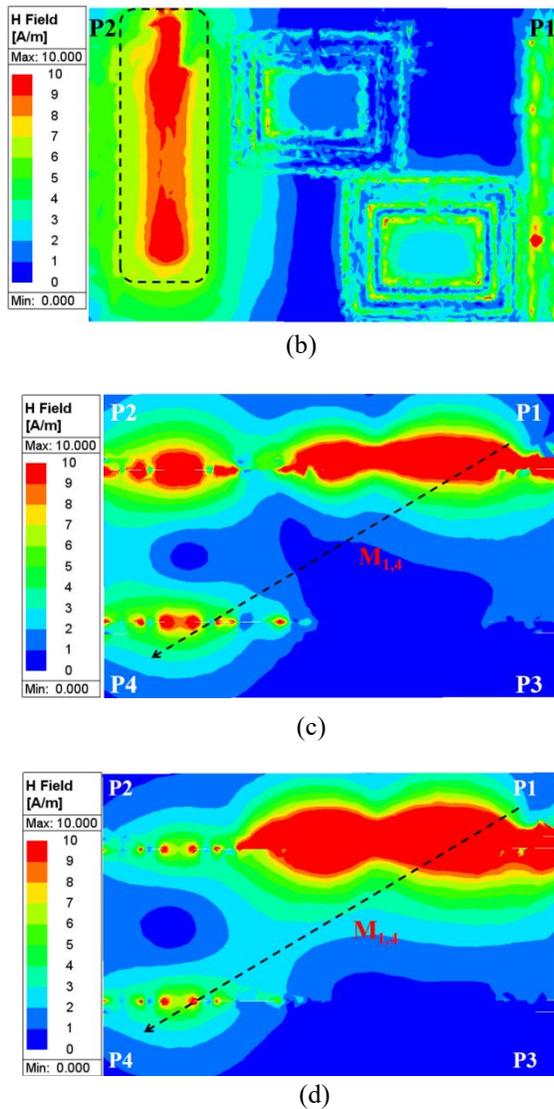

Fig.6. Simulated H filed of SWIPT system in the frequency of 40 MHz,(a) and (c): without absorber, (b) and (d) :with absorber. (a) and (b): the XY plane,(c) and (d): the XZ plane. $M_{1,4}$ means the mutual inductance of Port 1 and Port4.

Additionally, Fig.7 illustrates the simulated isolation performance of the system both without and with the proposed absorber. The isolation permance is evaluated by extracting the S-parameters, demonstrating how effectively the proposed absorber minimizes interference and isolation between the ports.The mutal inductance between Port1 and Port 4 is expected to be same as that between Port2 and Port3. The isolation performance is estimated by S parameter $|S_{1,4}|$. When the distance between Tx and Rx is 9mm, the estimated maximum isolation degradation is 5dB at a wireless power transfer(WPT) .In summary, the simulated results align closely with the theoretical predictions discussed earlier. For the proposed wireless power transfer(WPT) system, the optimal WPT frequency is identified where the real part of the permeability is approximately -1 and the imaginary part is close to 0. This specific condition ensures efficient power transfer and minimal magnetic losses, validating the design and effectiveness of the proposed absorber.

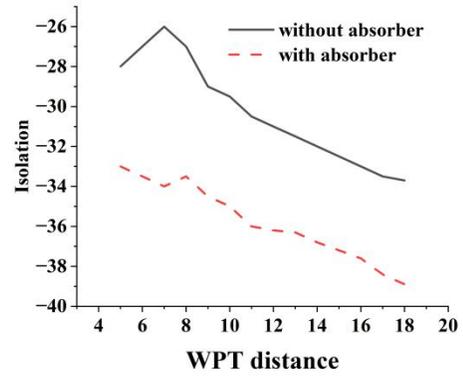

Fig.7. Isolation performance of the SWIPT system without and with absorber.

*D. Design and Simulation of proposed rectifier*

Fig.8(a) shows the SWIPT system integrated with proposed rectifier and Fig.8(b) depicts the model and equivalent circuit of rectifier.

A voltage doubler is used for proposed rectifier, the rectifier concludes two diodes and locates at Rx connected with Port 3. As illustrated in Fig.8(b), $C_1$ is connected with Port 3 and DC voltage is outputted by load $R_L$.

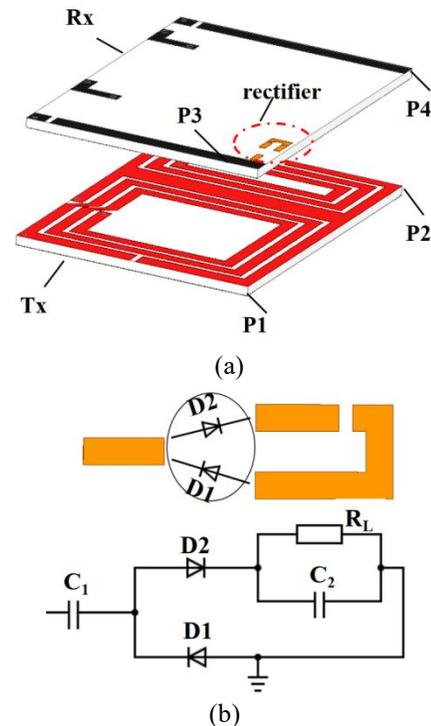

Fig. 8. (a)Proposed rectifier with SWIPT system.(b)Equivalent circuit and model of proposed rectifier.

Considering that the proposed rectifier needs to be connected to the SWIPT system at Rx power port, it is necessary to assess the impedance matching between them. In the simulation C1 and C2 is 2000pF and $R_L$ is 200Ω.

Specifically, the output impedance of the power port at Rx of the SWIPT system needs to be conjugately matched to the input impedance of the rectifier at 13.56MHz, as in

$$Zoutput\_Rx = Z*input\_rec \qquad (9)$$

Fig.9(a) shows the output impedance of Rx power port at 13.56MHz. As a result, to achieve conjugate matching with an output impedance of 48+3j Ω, in order to satisfy the conjugate impedance matching, the input impedance of proposed rectifier at 13.56MHz should be 48-3j Ω.

Fig.9(b) displays the input impedance performance of the rectifier with different input powers. It can be observed from the figure that when the input power is 24dBm, the input impedance of the rectifier is 48-3j Ω.

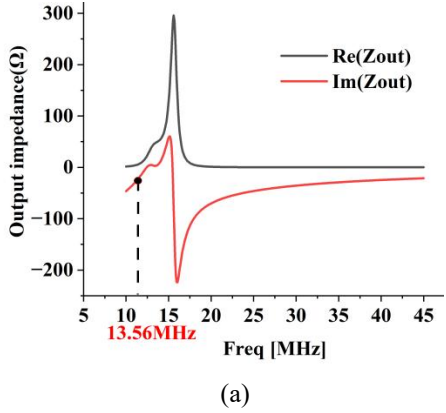

(a)

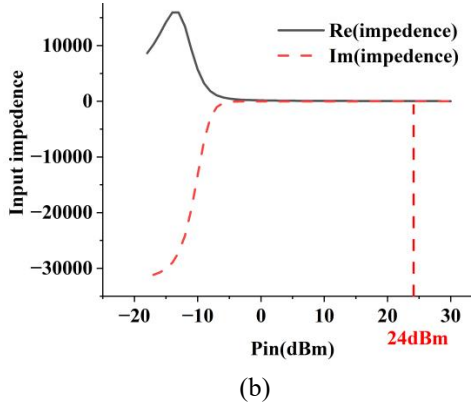

(b)

Fig.9.(a) Output impedance of proposed SWIPT system in Port 3.(b) Input impedance of proposed rectifier.

III. EXPERIMENT RESULTS

A. Setup of experiment

Fig.10(a) depicts the SWIPT system of Tx and proposed absorber, and Fig.10(b) shows Rx and rectifier (Type:HSMS-2860) respectively. Fig.10(c) illustrates the experiment sertup that the proposed rectifier is at the backside of Rx and the Rx is between two 5-mm human body equivalent phantom(X22017K01). Both of the top and bottom sides of the phantom are covered with a 0.1mm thick polythylene casing. Tx is placed 3.8mm above the phantom to ensure the distance of WPT system is 9mm, which can also eliminate the over-coupling effect. As same as in simulation used, the capacitance of both C1 and C2 is 2000pF and the impedance of load $R_L$ is 200Ω.

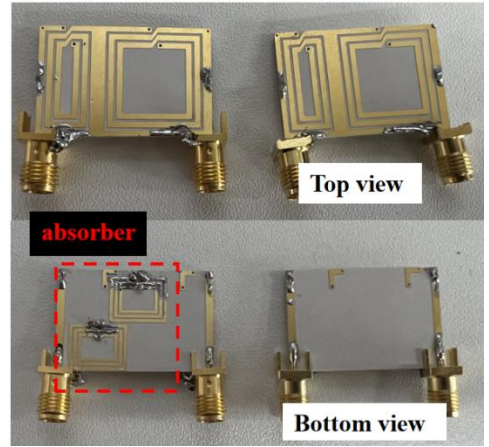

(a)

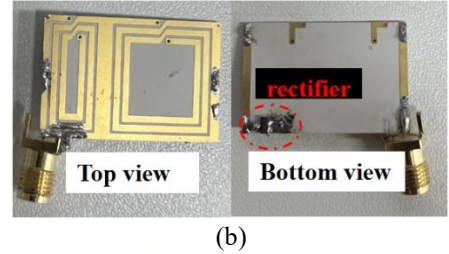

(b)

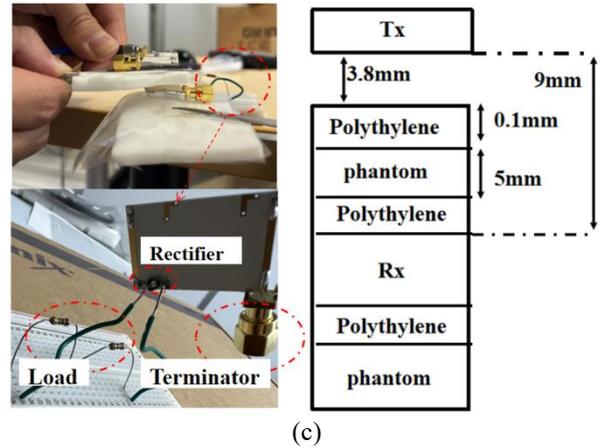

(c)

Fig.10.Fabricated components and experiment setup of the measurement.(a)Top view and bottom view of proposed Tx.(b)Top view and bottom view of proposed Rx.(c) Experiment setup.

B.S-parameter performance

Fig.11 illustrate the S-parameters of SWIPT systems with and without proposed absorbers. To achieve better impedance matching during measurements, 50-ohm terminators are employed on unconnected ports. A PNA analyzer (#N5222A) is used for these measurements. The $|S_{2,4}|^2$ [29] is measured to determine PTE(power transfer efficiency) of higher band while the isolation performance is measured using $|S_{1,4}|$. At a 9mm wireless power transfer(WPT ) distance, the isolation isolation performance is improve by approximately 5dB with the proposed absorber, reducing interference from P1. S11 and S22 means the reflection of Port 1(power port) and Port2(signal port).

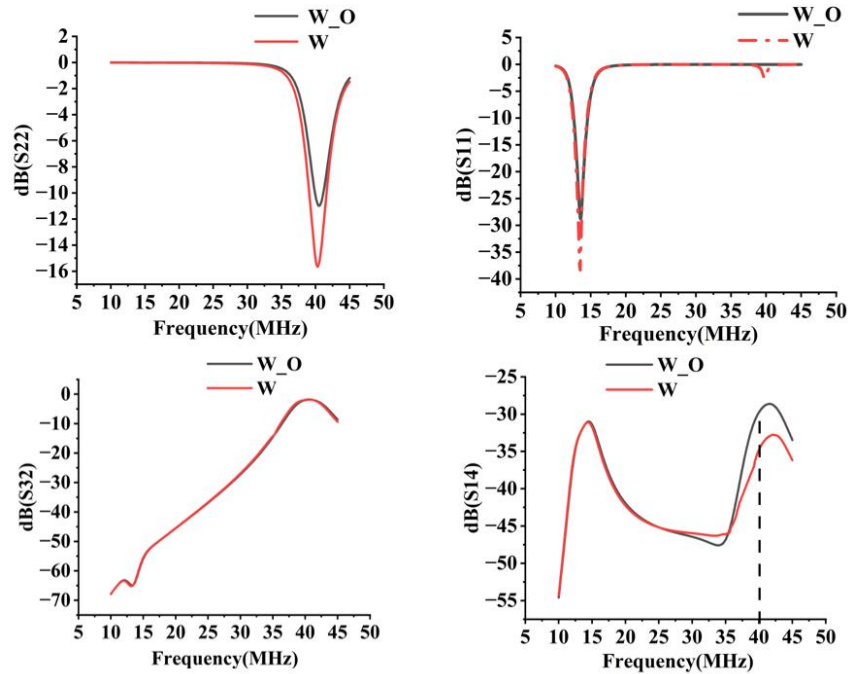

Fig.11.S-parameters of SWIPT system.Black line:without proposed absorber.Red line:with proposed absorber.(a).S22 of proposed SWIPT with and without absorber.(b).S11 of proposed SWIPT with and without absorber.(c).S32 of proposed SWIPT with and without absorber.(a).S14 of proposed SWIPT with and without absorberIn this proposed SWIPT system, the Port 3 is connected with rectifier so that there is no S-parameter of Port3.

*C. Measurement of BER and data rate*

Fig.12(a) depicts the configuration used for measuring bit error rate(BER). Two separate signal generator are utilized, one for delivering power and the other for providing signal bands. On the Tx side, the signal port generates and sends a pseudorandom bitstream, specifically a PRBS-9 sequence, modulated using minimum-shift keying(MSK) and the power port transmits a signal generated by series number MG3710A . On the Rx side, the power port is terminated with the rectifier. Concurrently, the signal port is linked to a signal and spectrum analyzer(series number:SMM100A).

In Fig.12(b) and (c), the data rate and BER results at the Rx are shown. The new absorber design achieves a substantial data rate increase of 7MHz/s, improving from 13.5MHz/s to 20.5MHz/s. Additionally, Fig.12(c) illustrates how the BER varies with the input signal power from Tx. The proposed SWIPT system shows stable communication with weaker signal power. Operating effectively between -23dBm to -29dBm, which is a 6dBm improvement over the original system.

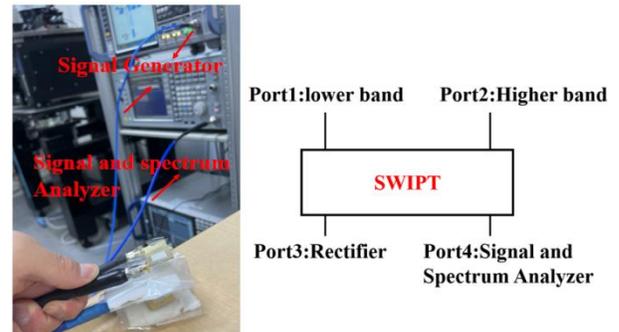

(a)

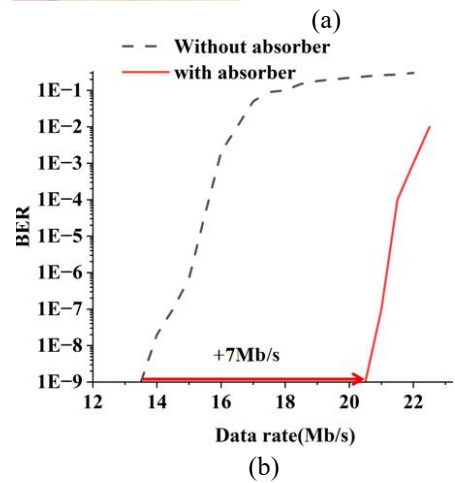

(b)

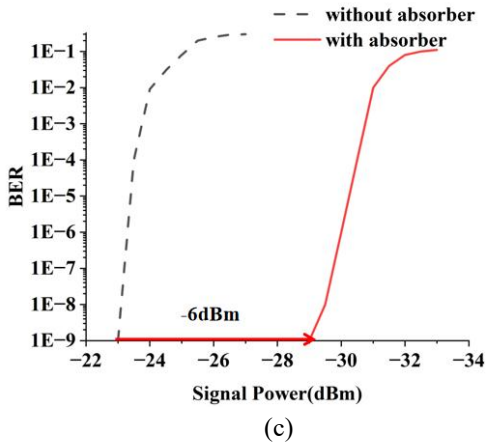

(c)

Fig.12.(a)Experiment setup for BER measurement and comparison of BER with (b)data rate.(c)power of input signal.BER=$10^{-9}$ with a condifence level of 99% when zero BER is observed.

*D. Measurement of PTE of SWIPT system*

Fig.13 shows the measurement of the output voltage of the proposed SWIPT system. The proposed absorber is at Tx side and the proposed rectifier is at Rx side. The output voltage of the load impedance of the rectifier is measured using a multimeter to determine the output voltage of the SWIPT system. Additionally, 50-ohm terminators are used to terminate the higher band in this measurement to make sure the best impedance matching.

The Tx and Rx components are fabricated by printing 0.017mm thick copper onto an Rogers RO3003 substrate. The substrate has a dielectric constant of 3, loss tangent of 0.001 and the thickness is 0.762mm. In this measurement, the input power is generated using vector signal generator(VSG, series number:SMM100A). Specifically, the lower band of Tx port is connected to the VSG. The input power of the SWIPT system is controlled by adjusting the output power of the VSG. Table II shows the output voltage measured by multimeter of the SWIPT system at different input powers with absorber and without absorber, respectively. In Table II, the output voltage of proposed SWIPT system is 5.12V when the input power is 24dBm. By using formula (10), the PTE of proposed SWIPT system can be obtained . The proposed system achieves RF-to-DC conversion 52% at a WPT distance of 9mm when the input power is 24dBm, at 13.56MHz.

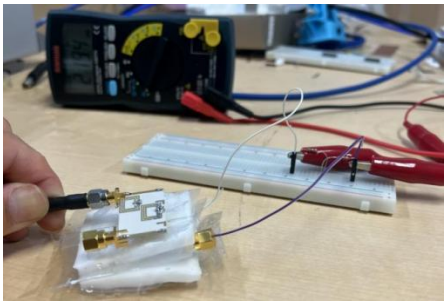

Fig.13.Measurement of the output voltage of the proposed SWIPT system.

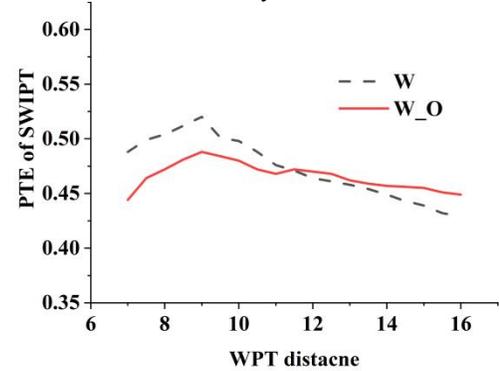

Fig.14 PTE of SWIPT system varies with WPT distance of the system without absorber and with absorber.

$$Eff\_system = (V\_output)^2 / (P\_input * R_L). \quad (10)$$

Fig.14 illustrates the PTE of SWIPT system both with absorber and without absorber, respectively. The X-axis is the distance between Tx and Rx. From Fig.14, with the help of proposed absorber, the efficiency of SWIPT system can be improved and the maximum of efficiency is 52% when the distance of WPT is 9mm. Also, because the Q factor of SRRs (Absorber unit) is low, the PTE decreased by almost 2% when the distance exceeds 12mm than the system without proposed absorber.

Table III provides a comparison of the performance of the proposed SWIPT system with recent solution designed for biomedical implants. Our system shows a marked improvement in data rate performance when compared to the solutions described in references[1],[5],[6].

Table II
OUTPUT VOLTAGE OF PROPOSED SWIPT SYSTEM

| Pin(dBm) | V_output(V)W_AB | V_output(V)WO_AB |
|---|---|---|
| -20 | 0 | 0 |
| -15 | 0 | 0 |
| -10 | 0 | 0 |
| -5 | 0.026 | 0.018 |
| 0 | 0.142 | 0.127 |
| 5 | 0.384 | 0.304 |
| 10 | 0.878 | 0.792 |
| 15 | 1.721 | 1.544 |
| 20 | 3.115 | 3.018 |
| 22 | 4.011 | 3.9 |
| 24 | 5.12 | 4.92 |

Table III

COMPARISON OF PROPOSED ABSORBER-BASED SWIPT WITH OTHER PAPERS WORK

| Work | Surroundings | Number of Ports | Frequency | SIZE(mm) | WPT distance | Rectifier | PTE | Data rate(Mb/s) |
|---|---|---|---|---|---|---|---|---|
| 1 | Chicken | 2 | 49MHz | TX:20×20 RX:10×10 | 10mm | NO | AC:57% | No |
| 5 | Air | 2 | 30MHz | 30×30 | 42mm | NO | AC:80% | 5 |
| 6 | Air | 2 | 35MHz | 30×30 | 45mm | NO | AC:63% | UP:4/Dn:2 |
| This work | Phantom | 2 | 13.56MHz/40MHz | 30×20 | 9mm | YES | DC:52% | without Absorber:13.5 with Absorber:20.5 |

## V. CONCLUSION

This study introduces a novel metamaterial-based absorber designed for a compact SWIPT system. Using CRLH TL theory and HFSS simulations, the absorber's permeability and H-field characteristics were analyzed to optimize isolation performance and minimize PTE degradation. We optimized the absorber unit cell to achieve a magnetic permeability near -1. We fabricated and tested SWIPT system both with and without the absorber using phantom sheets. Fabricated absorber were tested to measure S-parameters, BER, data rated and PTE. These results revealed that the absorber enhance isolation by 5dB at a 9mm WPT distance within the signal band, using a 5mm-thick human equivalent phantom. Notably, our proposed absorber demonstrated superior data rate performance and better tolerance of weaker signal power compared to previous studies. Moreover, the proposed rectifier which use conjugate matching method can achieve 52% RF-to-DC conversion efficiency in biological tissue.